\documentclass[a4paper,fleqn,usenatbib]{mnras}

\usepackage{txfonts}    

\usepackage[T1]{fontenc}
\usepackage{ae,aecompl}

\usepackage{graphicx}	
\usepackage{amssymb}	
\usepackage{ulem}	

\usepackage{bm}
\usepackage{natbib}

\title[Periods of Fast-Rotating Pulsating B-type stars]
      {Period-luminosity relations of fast-rotating B-type stars in the young open cluster NGC\,3766}

\author[H. Saio et al.]{ 
H. Saio,$^{1,2}$\thanks{E-mail: saio@astr.tohoku.ac.jp} 
S. Ekstr\"om,$^{2}$ 
N. Mowlavi,$^{2}$
C. Georgy,$^{2}$
S. Saesen,$^{2}$
P. Eggenberger,$^{2}$
\newauthor{
T. Semaan$^{2}$ and S. J. A. J. Salmon$^{3}$
}
\\
$^{1}$Astronomical Institute, Graduate School of Science, Tohoku University, Sendai 980-8578, Japan\\
$^{2}$Institute of Astronomy, University of Geneva, 51 chemin des Maillettes, 1290 Versoix, Switzerland\\
$^{3}$Institut d'Astrophysique et de G\'eophysique, Universit\'e de Li\`ege, 17 all\'ee du 6 Ao\^ut, 4000, Li\`ege, Belgium
}

\date{Accepted XXX. Received YYY; in original form ZZZ}

\pubyear{2016}

\begin{document}
\label{firstpage}
\pagerange{\pageref{firstpage}--\pageref{lastpage}}
\maketitle

\begin{abstract}
We study the pulsational properties of rapidly rotating main-sequence B-type stars using linear non-adiabatic analysis of non-radial low-frequency modes taking into account the effect of rotation. We compare the properties of prograde sectoral $g$ and retrograde $r$ modes excited by the $\kappa$ mechanism at the Fe opacity peak with the newly discovered period-luminosity relation that is obeyed by a group of fast-rotating B-type stars in the young open cluster NGC\,3766. The observed relation consists of two sequences in the period versus magnitude diagram, at periods shorter than 0.5 days. We find that this property is consistent with similar period-luminosity relations predicted for excited sectoral prograde $g$-modes of azimuthal orders $m=-1$ and $m=-2$ in fast-rotating stars along an isochrone. We further show that some of the rapidly rotating stars that have photometric variability with periods longer than a day may be caused by $r$-mode pulsation predicted to be excited in these stars. One fast-rotating star, in particular, shows both short and long periods that can be explained by the simultaneous excitation of $g$- and $r$-mode pulsations in models of fast-rotating stars. 
\end{abstract}




\section{Introduction}

A large fraction of the main-sequence band on the Hertzsprung-Russell (HR) diagram is covered by the presence of pulsating variable stars \citep[see e.g.,][]{jef16}, with a noticeable gap between the blue edge of $\delta$~Sct variables at $\log T_\mathrm{eff} \simeq 3.95$ and the red-edge of Slowly Pulsating B (SPB) stars at $\log T_\mathrm{eff} \simeq 4.05$.
The gap is understood, based on classical (i.e. non-rotating) stellar models, as resulting from the fact that the partial ionization zone of He (responsible for the $\kappa$ mechanism in $\delta$~Sct stars) is too close to the surface while the partial ionization zone of the iron peak elements (responsible for the $\kappa$ mechanism in SPB stars) is too deep inside the star.

Recently, \citet{mow13} discovered in the young open cluster NGC~3766 thirty six periodic variable stars that lie precisely in the gap region of the main sequence, at luminosities brighter than $\delta$~Sct stars and fainter than SPB stars.
The presence of periodic variable stars on the faint side of the SPB instability strip had already been sporadically reported in the literature, including Maia stars \citep[e.g.][]{sch98,per00}, individual late B-type stars \citep{kal04}, and some CoRoT targets \citep{deg09}, but the existence of such a population of periodic variables was most clearly revealed by \citet{mow13} from the study of an open cluster, in which the position of stars can be compared in the color-magnitude diagram to the positions of known groups of pulsating stars.
Similar data were also available for another young cluster, NGC~884 \citep{sae13}, though with a distribution of periods and stellar parameters that made conclusions more difficult to draw.
The results presented in \citet{mow13} triggered both theoretical investigations \citep[][ who suggested the new variables to be fast-rotating SPB stars]{sal14} 
and observational studies \citep[e.g.][]{lat14,lat16,bal16}. 

New key observational results have just been published by \citet{mow16} on the new periodic variables in NGC~3766, based on spectra obtained with the Very Large Telescope.
The authors first confirm the fast-rotating nature of all the new variables with periods less than 0.5~d.
Most importantly, their analysis reveals that the photometric periods of the majority of the fast-rotating stars fall on two ridges in the period-luminosity (PL) plane, revealing the existence of a new PL relation obeyed by those stars.
The relation, they argue, is linked to stellar rotation, with pulsation playing a key role in the formation of the sequences.
\footnote{Models with single or double surface spots for the light variations are inconsistent with the two ridges in the PL plane, because some stars have two  periods closely spaced (both belonging to one ridge), or have two separate periods corresponding to the two ridges, but the period ratio always deviates slightly from two.}

In this paper, we investigate the origin of the two PL sequences discovered for the Fast-Rotating Pulsating B-type (FaRPB) stars by \citet{mow16}, based on pulsation models of fast-rotating stars.
We consider stellar models lying on isochrones (Sect.~2), and analyze their pulsation properties for both $g$ and $r$~modes using a linear non-adiabatic prescription (Sect.~3). 
Our pulsation predictions are then compared to the data of FaRPB stars
in NGC~3766 (Sect.~4).
Conclusions on the nature of the PL relation of FaRPB stars and on the occurence of $r$~mode pulsation in some periodic stars observed in the cluster are finally drawn in Sect.~5.
In Appendixes~\ref{sec:lambda} and \ref{sec:growth} we discuss the general properties, mainly in the co-rotating frame, of low-frequency pulsations in a rapidly rotating star. Finally, Appendix~\ref{sec:opalop} compares SPB  instability regions based on the OPAL and OP opacities with slightly different metallicity.

\section{Models}

\subsection{Evolution models}

We compute models with rotation using the Geneva stellar evolution code \citep[see][]{eks12,and16}.
All models start at the zero-age main sequence (ZAMS) with an initial homogeneous chemical composition of X=0.72 and Z=0.014 needed to reproduce the composition of the Sun at its present age when we evolve a 1~M$_\odot$ model up to the present age of the Sun \citep[see][]{eks12}.
We also compute a few $6\,M_\odot$ models with an initial larger-than-solar metallicity of Z=0.02 and with X=0.706, that are used for comparison purposes. 
The opacities are taken from the OPAL opacity tables \citep{igl96}, adapted to the chemical compositions of our models.

Several sets of evolutionary tracks are computed, each starting with different masses and (uniform) rotation rates; 
the latter ones being expressed in terms of the  $\Omega_\mathrm{init}/\Omega_\mathrm{c}$
ratio of the angular velocity to critical angular velocity. 
With the critical angular frequency, the centrifugal force is equal to the gravity at the equator \citep[see][]{mae00}.
The stellar masses used in this study range from 2.5 to 8~$M_\odot$, and the initial $\Omega_\mathrm{init}/\Omega_\mathrm{c}$  
ratios from 0 and 0.8.
To be consistent with the pulsation analysis, which assumes solid-body rotation, we artificially added a large diffusion coefficient ($D_\mathrm{arti.} = 10^{15}\,\mathrm{cm}^2\,\mathrm{s}^{-1}$) in the angular momentum transport equation \citep[see][]{cha92}. 
For models reaching the critical velocity in the course of their evolution, a mechanical mass-loss is assumed to keep the star at critical velocity \citep{geo11}.

\subsection{Nonradial pulsations of rotating stars}

The latitudinal and azimuthal dependence of the amplitude of linear nonradial pulsations in a non-rotating star (or rotating with a period much longer than the pulsation period) is separated by spherical harmonic $Y_\ell^m(\theta,\phi)$ from the radial distribution, where $\theta$ and $\phi$ are the colatitude and azimuthal angles, respectively. 
The pulsation motion ${\bm \xi}({\bm r},t)$ and perturbations $\delta f({\bm r},t)$ of scalar variables associated with the nonradial mode identified with a latitudinal degree $\ell$,  an azimuthal order $m$ and  a radial order $n$, can be represented as
\begin{eqnarray}
  {\bm\xi}({\bm r},t) & = & e^{i\omega t}\!\!\left(\xi_{r,n} Y_\ell^m {\bm e}_r + \xi_{{\rm h},n}\nabla_{\rm h}Y_\ell^m\right), \nonumber\\
  \delta\! f({\bm r},t) & = & e^{i\omega t}\delta\! f_{n} Y_\ell^m,
\label{eq:norot}
\end{eqnarray}
where $\omega$ is the angular frequency of pulsation, and $\nabla_{\rm h}$ is the horizontal gradient defined as
\begin{equation}
  \nabla_{\rm h}Y_\ell^m = {\partial Y_\ell^m\over\partial\theta}{\bm e}_\theta + {1\over\sin\theta}{\partial Y_\ell^m\over\partial\phi} {\bm e}_\phi.
\label{eq:nabla}
\end{equation}
In addition to the displacement form given in equation\,(\ref{eq:norot}) (we call it a spheroidal motion), toroidal motions expressed as
\begin{equation}
{\bm\eta}\propto(\nabla_{\rm h}Y_{\ell'}^m){\bm\times}{\bm e}_r
\label{eq:toroid}
\end{equation}
are possible in general.  In a non-rotating star they generate no oscillation and do not couple with spheroidal motions.
As a result, toroidal motions have no effect on nonradial pulsations in non-rotating stars. 

In the presence of rotation, however, the Coriolis force connects the spheroidal motion associated with $Y_\ell^m$ with the toroidal motions associated with $Y_{\ell\pm1}^m$; i.e., $\ell' = \ell \pm 1$, which brings coupling among spheroidal  motions of $\ell, \ell \pm 2, \ldots$~. 
Therefore, in the presence of rotation, the angular dependence of a nonradial pulsation cannot be represented by a single spherical harmonic; i.e., a single $\ell$ cannot be specified to a pulsation mode.   
We note, however, that the even or odd property of a mode with respect to the equator (which is determined by even or odd $\ell-|m|$) is preserved because the coupling occurs between $\ell$ and $\ell\pm 2$.  

Since the Coriolis force is proportional to $|2\Omega\omega|$ and pulsational acceleration is proportional to $\omega^2$, the Coriolis force effects are stronger for larger $2\Omega/\omega$, where $\omega$ is the pulsation frequency in the co-rotating frame and $\Omega$ is the rotation frequency.
Therefore, low frequency $g$ modes are affected by the Coriolis force more strongly than $p$ modes.

Furthermore, stellar rotation makes possible the presence of pulsation modes in which toroidal motions are dominant.
These modes are called $r$ modes \citep{pap78}.
They are normal modes of global Rossby waves, whose restoring force is associated with the latitudinal gradient of the Coriolis force.
They propagate in the opposite direction of rotation; i.e., they are retrograde modes in the co-rotating frame.
Since the Coriolis force on the toroidal motion generates spheroidal perturbations, and hence density perturbations, normal modes ($r$ modes) are formed \citep{pap78, pro81} and can be excited by the $\kappa$ mechanism \citep{sai82,ber83} which works on the spheroidal perturbations.  
In the co-rotating frame, the frequency range of the $r$ modes lie in the frequency range of high-order $g$ modes.
In a numerical analysis, a group of $r$ modes is recognized as a series of frequencies with small radial orders ($n$) leading to a maximum frequency ($n=1$) in a middle of the $g$-mode frequency range.

In this paper, we perform a linear nonadiabatic analysis of the $g$ and $r$ modes by expressing the pulsational displacements ${\bm \xi}$ and the perturbations $\delta\! f$ of the scalar variables of a mode as truncated series \citep{lee95} in the form of
\begin{eqnarray}
  {\bm\xi} & = & e^{i\omega t}\sum_{j=1}^J\left[\xi_r^j Y_{l_j}^m {\bm e}_r + \xi_{\rm h}^j\nabla_{\rm h}Y_{l_j}^m + \eta^j(\nabla_{\rm h}Y_{l'_j}^m){\bm\times}{\bm e}_r \right] ,
  \label{eq:expand1}
  \\
  \delta\! f & = & e^{i\omega t}\sum_{j=1}^J\delta\! f^jY_{l_j}^m ,
\label{eq:expand2}
\end{eqnarray}
where $l_j = |m| + 2(j - 1) + I$ and $l'_j = l_j +1 - 2I$ with $I=0$ for even modes and $I=1$ for odd modes,   and $J$ is the truncated length of the series. We adopt $J=8$ in most cases.
Although no mode in a rotating star can be described by a single latitudinal degree, we sometimes use, for convenience,  $\ell_{\rm e}$ or $\ell_{\rm e}'$ to identify a mode, representing 
the main spheroidal degree or main toroidal degree, respectively. 
We note that a mode has an azimuthal order $m$, whose sign indicates the azimuthal propagation direction of the mode in the co-rotating frame. In this paper, we adopt the convention that $m < 0 $ corresponds to a prograde mode, which is consistent with the above equations.
The amplitude is expected to grow if the imaginary part of $\omega$ is negative with a growth rate of $-{\cal I}(\omega)/{\cal R}(\omega)$, where ${\cal I}$ and ${\cal R}$ mean the imaginary and the real parts. 
If a mode has a positive growth rate, we call it an excited mode.
Because of the linear analysis, we cannot predict the amplitude of the excited mode.

To avoid the complexity related to the inseparability of the angular dependence of the equations, the traditional approximation of rotation (TAR) is sometimes used in the literature, in which the horizontal component of the angular rotation frequency $\Omega\sin\theta$ is neglected.
This corresponds to neglecting the Coriolis force associated with the radial displacement, and the radial component of the Coriolis force associated with the horizontal displacement in the momentum equation.
In this approximation the angular dependence of a pulsation mode is separated from the radial dependence, and the set of equations becomes similar to the one for the non-rotating case.
The TAR gives reasonable frequencies for $g$ and $r$ modes, and is useful in understanding the properties of low-frequency modes in a rotating stars (see  Appendix~\ref{sec:lambda}).
However, the stability of retrograde $g$ modes as well as tesseral $g$ modes is significantly affected by the approximation, because the effect of mode interactions are not included in the TAR  
\citep{lee08}.
We do not use the TAR in this paper, because evaluating accurately the stability of each mode is important for the present study.

In rapidly rotating stars, retrograde $g$ modes and tesseral $g$ modes tend to be damped \citep[][]{lee08,apr11}, while sectoral prograde $g$ modes are less affected by rotation. 
In our models for rapidly rotating B-type main-sequence stars of intermediate mass stars, all excited retrograde modes are found to be $r$ modes. Although some tesseral odd prograde and axisymmetric modes are excited, they have large latitudinal wavenumber, which corresponds to the fact that the terms of large $l_j$ ($> |m| +1$) in eqs.~\ref{eq:expand1},~\ref{eq:expand2} are important (or, in words of the TAR, $\lambda > \ell(\ell+1)$; Appendix~\ref{sec:lambda}). The amplitude distribution on the surface is confined to an equatorial region \citep[][; Appendix~\ref{sec:growth}]{tow03} and anti-symmetric to the equator, so that they should be hardly visible because of the cancellation for a star with a large $v \sin i$. 
On the other hand, the latitudinal distribution of a prograde sectroral $g$ mode is affected only mildly by rotation \citep[][Appendix~\ref{sec:growth}]{lee97,tow03} so that they should be most visible in a star having a large $v \sin i$. 
The amplitude distribution of an $r$ mode tends to have a broad peak in the mid-latitude \citep[][ Appendix~\ref{sec:growth}]{lee97,sav05}, so that $r$ modes can be detected photometrically if they produce enough temperature variations. 
The theoretical prediction is consistent with the period spacings observed by \citet{vanr16} in the Kepler data of $\gamma$ Dor variables.
These authors found that most of the prograde modes detected in rapidly rotating $\gamma$ Dor stars are prograde sectoral $g$ modes, while all retrograde modes are $r$ modes.
Taking into account these facts, we analyse in this paper the behaviour of both sectoral prograde $g$ modes and $r$ modes in intermediate-mass main-sequence stars.

\begin{figure}
	\includegraphics[width=\columnwidth]{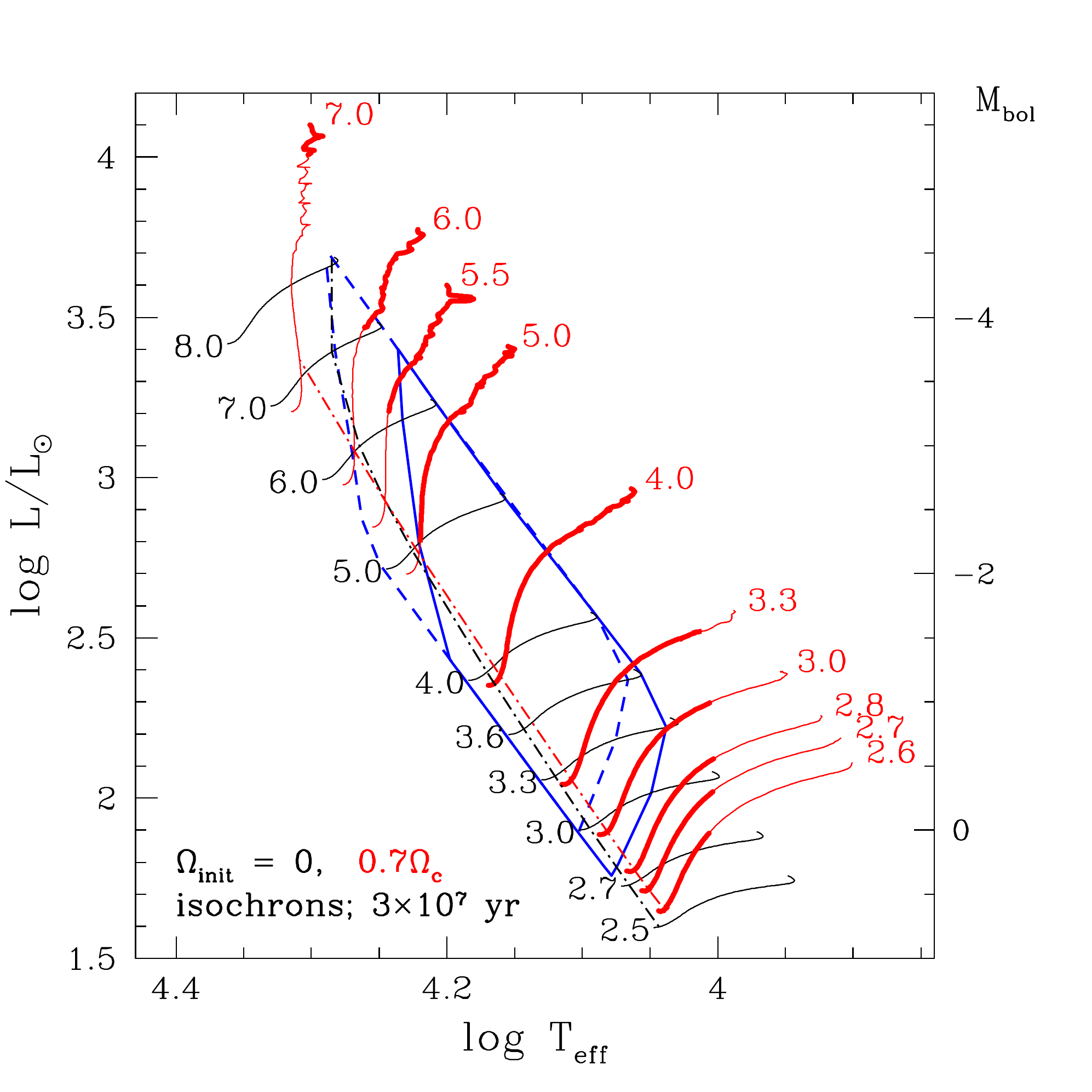} 
        \caption{Theoretical HR diagram showing evolutionary tracks of non-rotating (black lines) and rotating (red lines) models assuming an initial rotation of 70\% the critical angular velocity. A chemical composition of $(X,Z) = (0.72, 0.014)$ is adopted. The numbers close to the starting or end point of each evolutionary track indicate the initial masses in solar unit. The black and red dot-dashed lines are isochrones at $3\times10^7$yr for non-rotating and rotating models, respectively. The solid and dashed lines in blue are the instability boundaries for $\ell=1$ and $\ell=2$ $g$ modes, respectively, for non-rotating models. The thick parts of the evolutionary tracks of rotating models indicate where prograde sectoral dipole modes ($m = -1$) are excited.
    }
    \label{fig:hrd}
\end{figure}

\section{Excitation of $g$ and $r$ modes in rotating main-sequence B-type stars}
\label{Sect:grModesExcitation}

High order $g$~modes are excited in intermediate-mass main-sequence B-type stars by the $\kappa$-mechanism of the opacity peak at a temperature of $\sim$2$\times10^5$~K \citep{gau93,dzi93}.   
The blue lines in Fig.~\ref{fig:hrd} indicate the region in the HR diagram where high-order $g$ modes are excited in non-rotating models (solid lines for $\ell=1$, dashed lines for $\ell=2$ modes).
Stars located in this region of the HR diagram show multi-periodic light variations with periods from one to a few days, characteristic of $g$ modes. They are SPB  stars. 

The cool-luminous boundary of the SPB instability range arises due to strong radiative damping in the radiative core at the termination of the main-sequence evolution; i.e.,  the boundary indicates the disappearance of the convective core. 
Since rapidly rotating stars have a more extended main sequence in the HR diagram than non-rotating stars due to rotational mixing in the stellar interior \citep[e.g.,][]{mae09}, the instability region of $g$ modes is accordingly wider for models of rotating stars than for models of non-rotating stars.
This is shown, in Fig.~\ref{fig:hrd}, by thick-line parts of evolutionary tracks for rotating stars 
starting with an initial angular velocity equal to 70\% the critical angular velocity. 

In addition, the lower luminosity limit of the SPB instability range is predicted to be lower (i.e., occurs in  less massive stars) in models of rapidly rotating stars than in models of non-rotating stars, as first found by \citet{tow05a}.
\citet{sal14} nicely explained it by combining the optimal condition of the $\kappa$-mechanism with an increase of the periods (in the co-rotating frame) of sectoral prograde modes due to rotation.

\subsection{Periods and period-spacings}

\begin{figure*}
   \includegraphics[width=\columnwidth]{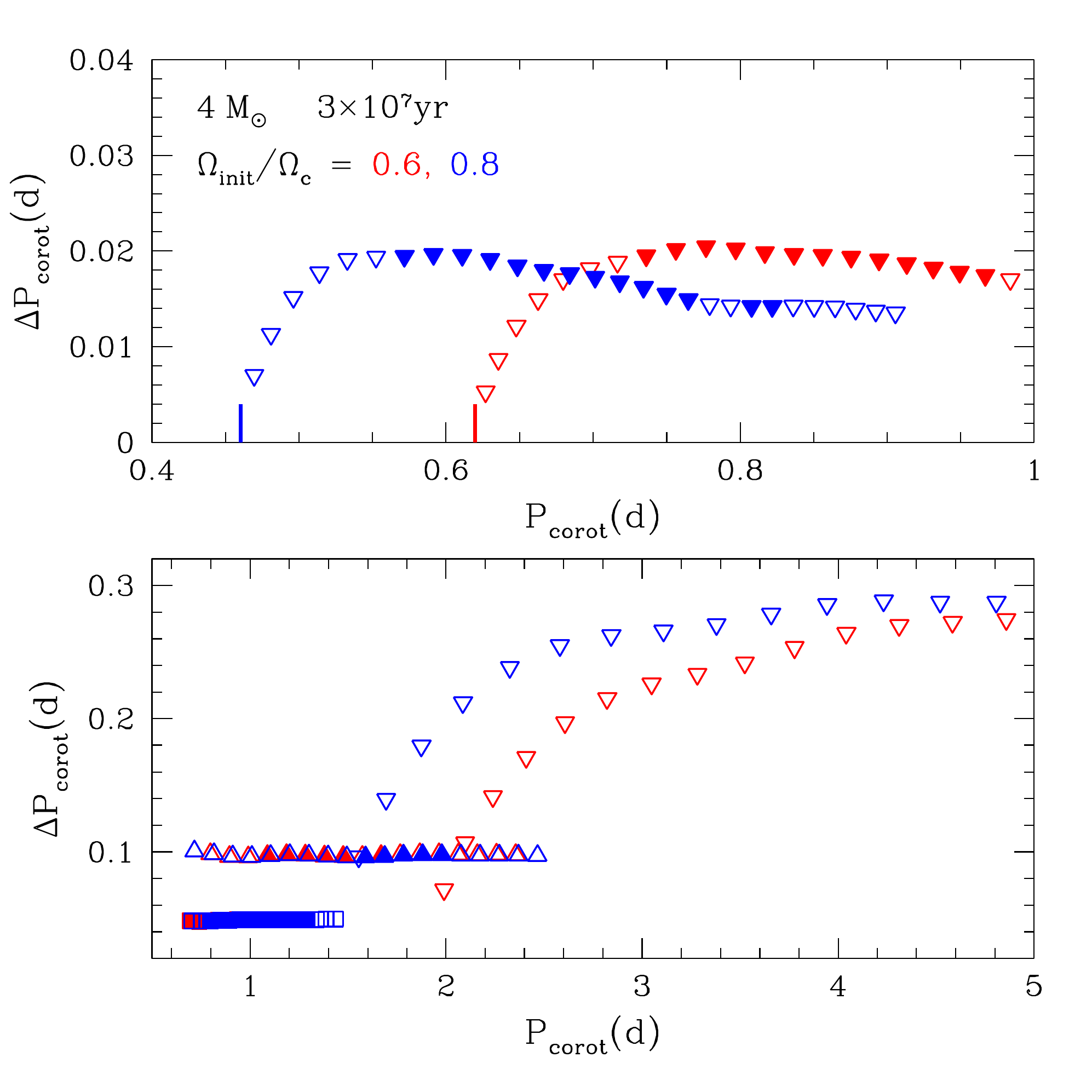}  
   \includegraphics[width=\columnwidth]{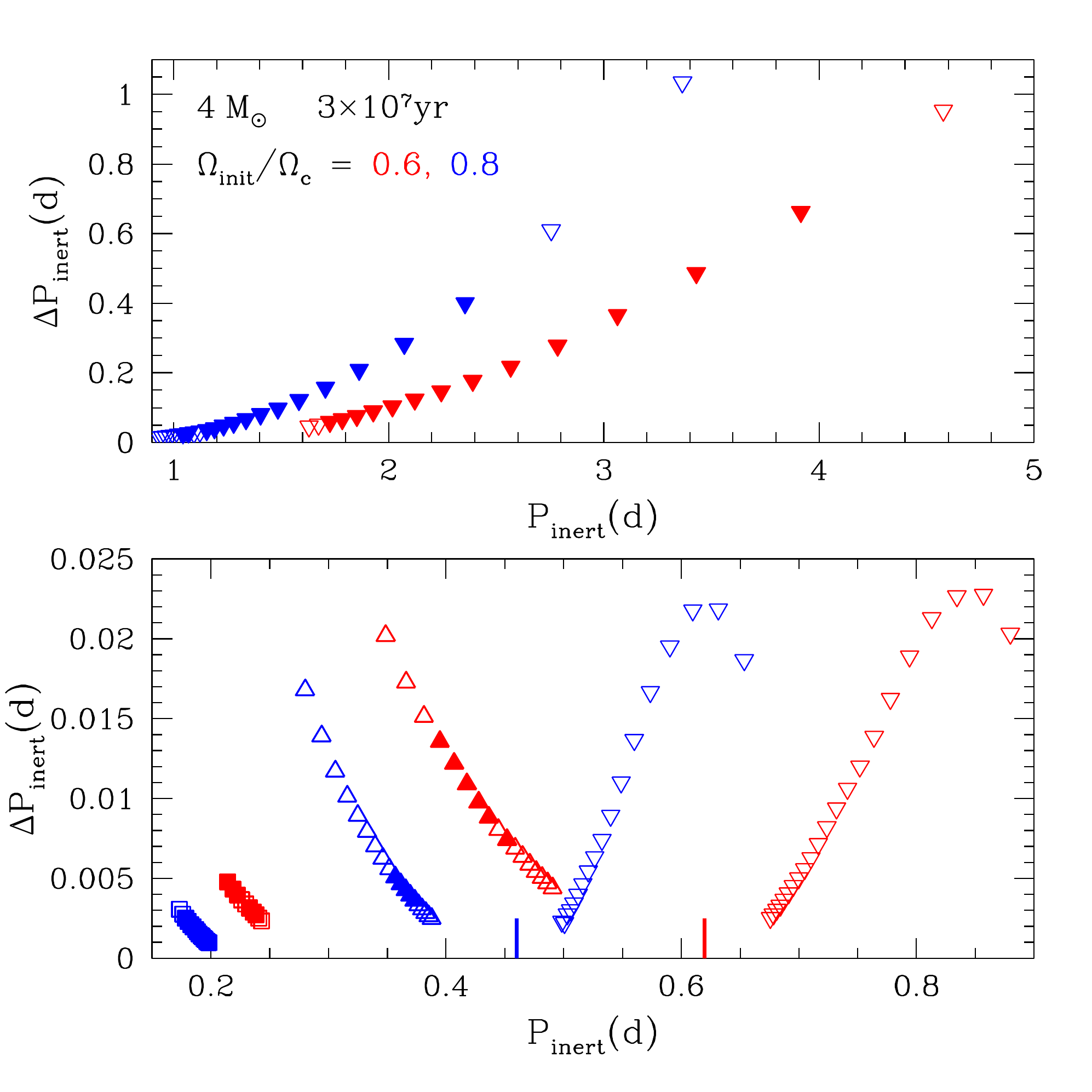}  
   \caption{Period-spacings  versus periods for $g$ and $r$~modes of a rotating $4M_\odot$ main-sequence model at an age of $3\times10^7$yr in the co-rotating frame (left panels) and in the inertial frame (right panels).  Filled and open symbols are for excited and damped modes, respectively. Initial (i.e. at ZAMS) angular rotation velocities are color-coded as indicated in the top panels. Triangles and squares are prograde sectoral $g$~modes of $m = -1$ and $m=-2$, respectively, while inverted triangles are $r$~modes of $m = 1$ with $\ell'_{\rm e} = 1, \ell_{\rm e} =2$ (upper panels) and with $\ell'_{\rm e} = 2, \ell_{\rm e}=1$ (lower panels). Lower and upper panels are, respectively, for modes having even and odd temperature distributions  with respect to the equator.   The color-corded short vertical lines drawn on the bottom horizontal axis indicate the rotation periods. 
   }
   \label{fig:space}
\end{figure*}

It is known that the periods of high radial order 
$g$ modes of non-rotating stars are equally spaced because the periods are approximately proportional to $n/\sqrt{\ell(\ell+1)}$ \citep[e.g.,][]{unno,ack10}.
This property of equal period spacing is preserved for high-order prograde sectoral $g$ modes in rapidly rotating stars if the periods $P_{\rm corot}$ in the co-rotating frame are longer than the rotation period $P_{\rm rot}$.
This is shown by the triangles and squares in the bottom-left panel of Fig.\,\ref{fig:space}.
We note that the period spacing $\Delta P_{\rm corot}$ is larger for rotating stars than for non-rotating stars because of a decrease in the effective $\ell(\ell+1)$ ($=$ eigenvalue $\lambda$ in the TAR; see Appendix~\ref{sec:lambda}).
 
The period spacing properties of $g$ modes in Fig.\,\ref{fig:space} look very different from that of sectoral prograde dipole $g$ modes presented in fig.\,4 of \citet{bou13}, in which $\Delta P_{\rm corot}$  gradually increases with many cyclic variations as $P_{\rm corot}$ increases.
The cyclic variations are caused by the gradient of the mean-molecular weight ($\mu$) exterior to the convective core boundary \citep{mig08}.  No cyclic variations occur for our rotating models at an age of $3\times 10^7$\,yr, because no strong $\mu$ gradients are formed due to their young age and 
to the action of rotational mixing.
The gradual increase of $\Delta P_{\rm corot}$ is caused by the gradual decrease of  $\lambda$ as a function of the ratio $P_{\rm corot}/P_{\rm rot}$. The decrease of $\lambda$ saturates for $P_{\rm corot} > P_{\rm rot}$ (Fig.\,\ref{fig:lambda} in Appendix~\ref{sec:lambda}),  
so does the increase of $\Delta P_{\rm corot}$.  The saturation can be seen in fig.\,4 of \citet{bou13} to occur for $P_{\rm corot} \ga 1.8$\,d ($\approx P_{\rm rot}$) in the most rapidly rotating case. 
Therefore, the period spacings we find for prograde sectoral $g$ modes in Fig.\,\ref{fig:space} are consistent with those of \citet{bou13}.

The period in the inertial (observer's) frame, $P_{\rm inert}$, is related to the period in the co-rotating frame as
\begin{equation}
P_{\rm inert} = \left|{1\over P_{\rm corot}} - m{\Omega\over2\pi}\right|^{-1} .
\label{eq:pinert}
\end{equation}
Because of the rotational `advection' effect, $P_{\rm inert}$ of a prograde $g$~mode is shorter than $P_{\rm corot}$. 
As seen in the bottom panels of Fig.\,\ref{fig:space}, the period spacings of prograde $g$ modes   in the inertial frame are much smaller than those in the co-rotating frame, and decrease with period.  This  property is useful in asteroseismic analyses. 
In fact, such a period-spacing/period relation was recently observed in the Kepler data of a SPB star \citep{pap15} and used to infer the interior rotation rate and overshooting efficiency \citep{mor16}. 

In addition to $g$ modes,  some $r$ modes (odd modes, though) are excited in rotating stars in the SPB range, as shown by \citet{tow05} and \citet{sav05} using the TAR, and by \citet{lee06} without using the TAR.
Our calculations confirm that odd $r$ modes of $m=1$ ($\ell_{\rm e}' = 1, \ell_{\rm e} = 2$) are excited in models of rotating stars, as shown in the upper panels of Fig.\,\ref{fig:space}  (filled inverted triangles; see also Fig.~\ref{fig:growth} in Appendix~\ref{sec:growth}), while all even $r$ modes of $m=1$ ($\ell_{\rm e}'=2, \ell_{\rm e} = 1$; lower panels) seem to be damped.

In the co-rotating frame, each $r$-mode series has a limiting (maximum) angular frequency of
\begin{equation}
\omega_0 = {2m\Omega\over \ell_{\rm e}'(\ell_{\rm e}'+1)}
\label{eq:rmodf}
\end{equation}
(see Appendix~\ref{sec:lambda}).
The lowest radial order mode has a frequency close to this frequency, while modes of higher radial orders have smaller frequencies in the series \citep[e.g.,][]{pro81,sai82}. 
In other words, the frequencies of the $r$ modes in the co-rotating frame are confined between $0$ and $\omega_0$. Period spacings of sufficiently high order modes are nearly constant in the co-rotating frame (left panels in Fig.\,\ref{fig:space}).

For the $m=1$ odd $r$ modes ($\ell_{\rm e}'=1$), $\omega_0 = \Omega$ (phase speed equals rotation speed, but in the opposite direction). The limiting frequency in the inertial frame is therefore zero; i.e., $P_{\rm rot} < P_{\rm inert} < \infty$. 
As a result, for this series, the period spacing and the period in the inertial frame increase with decreasing radial order, as seen in the upper right panel of Fig.\,\ref{fig:space}. 
We note that $r$~modes with periods of a few days in the inertial frame are excited, allowing the co-existence of short-periods prograde $g$  modes and long-period $r$ modes in a rapidly rotating star. 

For even $r$ modes with $(m,\ell_{\rm e}, \ell'_{\rm e})=(1,1,2)$, $\omega_0={1\over3}\Omega$.  This corresponds to a limiting frequency of ${2\over3}\Omega$ in the inertial frame; i.e, the period range of this $r$ mode series in the inertial frame satisfies $P_{\rm rot} < P_{\rm inert} < 1.5\,P_{\rm rot}$. 

Recently, \citet{vanr16} identified retrograde modes in $\gamma$ Dor stars as $r$ modes based on the period/period-spacing diagram. 
Those $r$ modes 
should be even modes of $m=1$ judging from Fig.\,9 of \citet{vanr16}.
In our intermediate-mass main-sequence stars, however, only odd modes are excited among the $r$ modes by the $\kappa$-mechanism on the Fe opacity peak.  This is in agreement with the results obtained by \citet{lee06}.

\subsection{Period-luminosity relations along isochrones}

\begin{figure*}
   \includegraphics[width=\columnwidth]{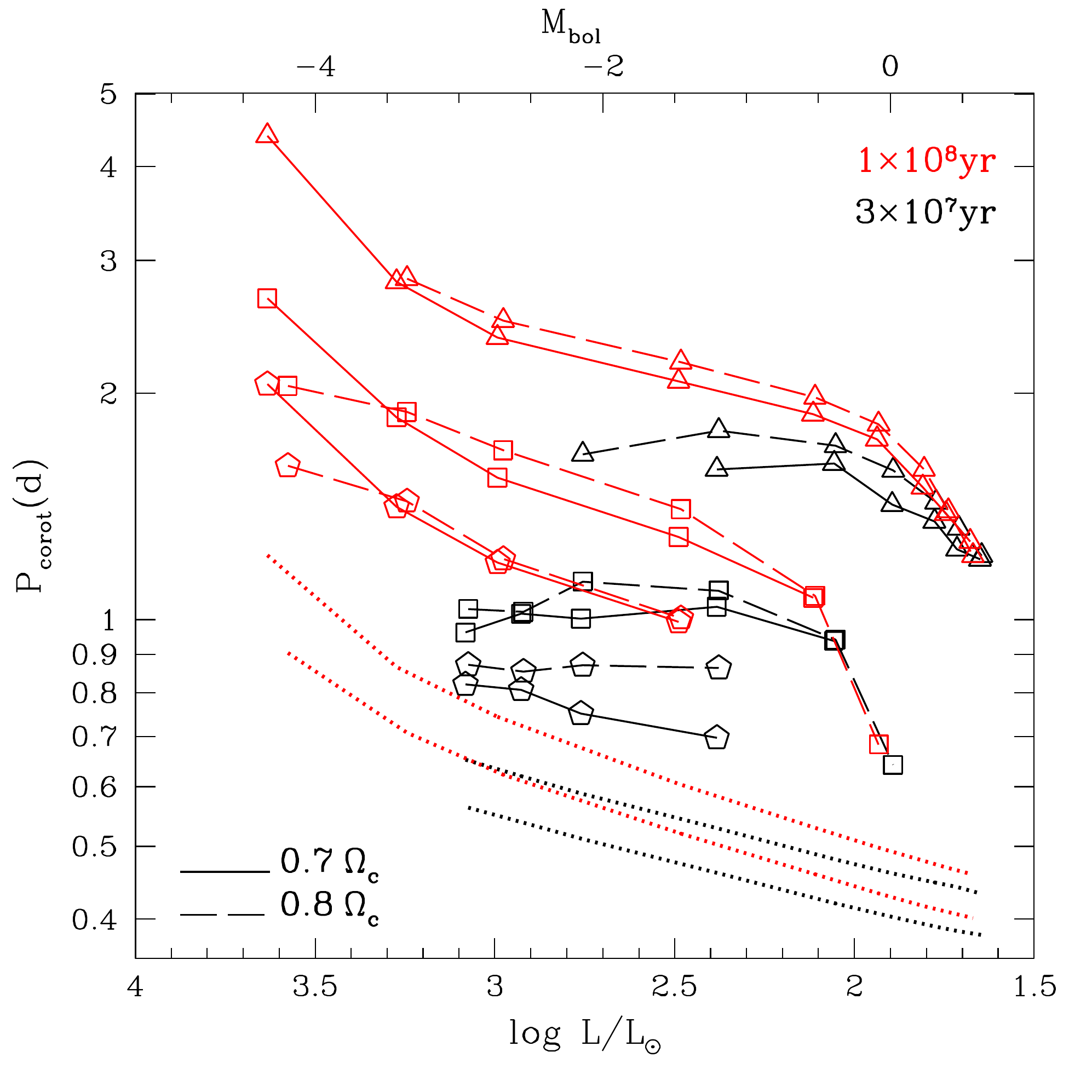}    
   \includegraphics[width=\columnwidth]{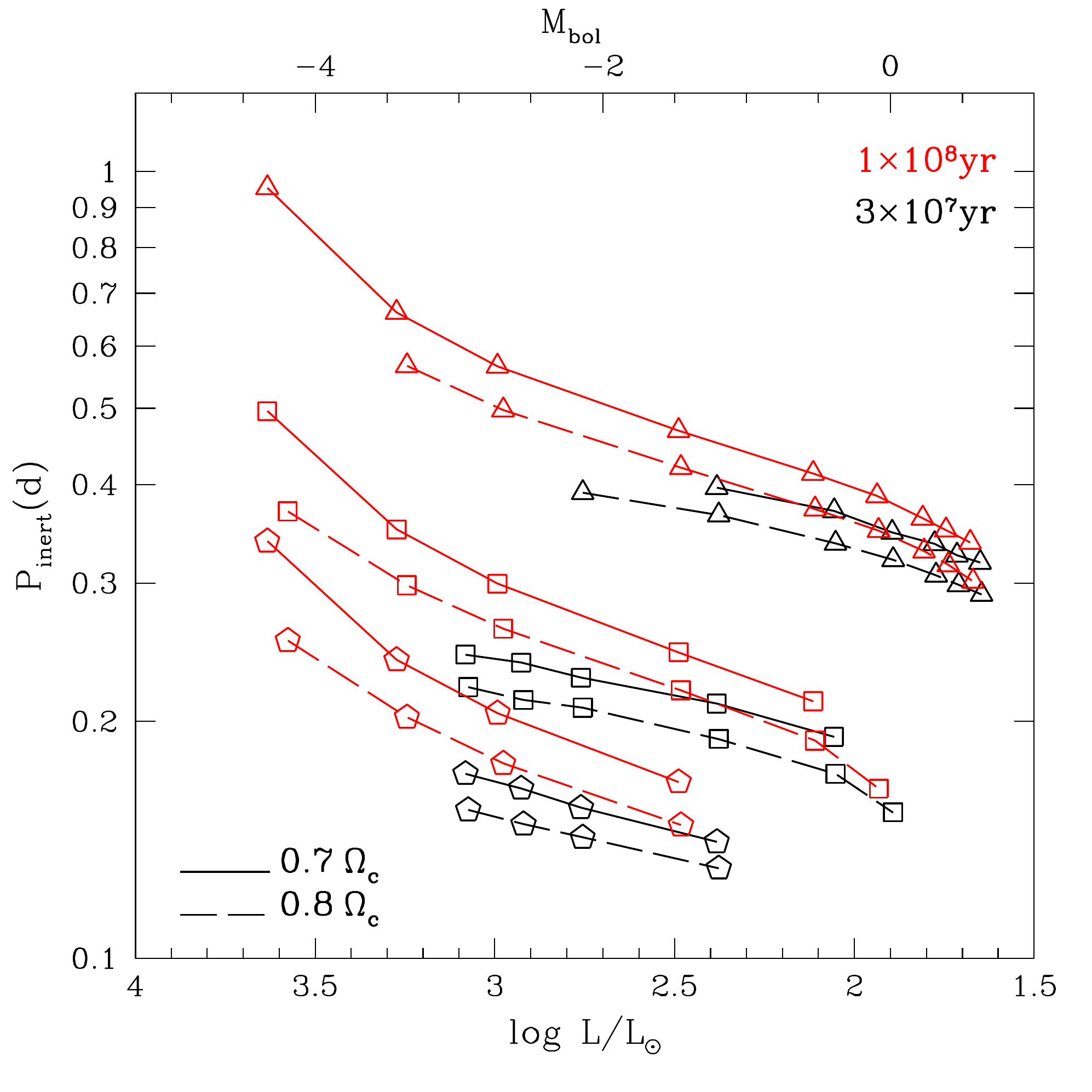}    
   \caption{Periods in the co-rotating frame (left panel) and in the inertial frame (right panel) of the most strongly excited modes (with highest growth rates) in the models at an age of 100 million (in red) and of 30 million (in black) years versus the luminosity of the models. Triangles, squares, and pentagons are prograde sectoral modes of $m=-1$, $m=-2$, and $m=-3$, respectively (with $m=-\ell_{\rm e}$). Models with initial rotation frequencies of $0.7\,\Omega_{\rm c}$ and $0.8\,\Omega_{\rm c}$ are connected by solid and dashed lines, respectively.
Dotted lines in the left panel show the rotation period as a function of luminosity along the isochrones for the two cases of the initial angular velocity of rotation. 
   }
   \label{fig:PL}
\end{figure*}

In each model,  prograde $g$ modes are excited only in a limited range of periods, which are, in the co-rotating frame,  considerably longer than the rotation period divided by $|m|$ (see the left panel of Fig.~\ref{fig:PL}). 
For this reason, the periods of the excited sectoral prograde $g$ modes  in the inertial frame are close to the rotation period divided by $|m|$ (see Eq.\,\ref{eq:pinert})
\footnote{Such groups of $g$ modes are observed in rapidly rotating Be stars \citep[e.g.,][]{wal05,cam08}.}.
Therefore, we expect a PL relation for each $m$ along an isochrone if the rotation frequency varies in a systematic way as a function of luminosity.
The right panel of Fig.\,\ref{fig:PL} shows such PL relations for ages of $3\times10^7$ and $1\times10^8$\,yr, in which the periods of the most strongly excited mode (with the highest growth rate) are connected for $\ell_{\rm e} (= -m) = 1$ (triangles), $\ell_{\rm e}=2$ (squares),  and $\ell_{\rm e}=3$ (pentagons). The
$g$ modes of each $m$ form a distinctive PL relation for a given initial angular velocity to critical angular velocity ratio.
As seen in this figure, the gradients of the PL relations at a given age are not sensitive to the initial rotation velocity, while the period of the most excited mode at a luminosity shifts upwards as rotation speed decreases.
For this reason, we expect 
a  tight PL relation to be observed in a young cluster, where we expect that stars were born with large rotation speeds not far from the critical speeds because of the matter accretion in the formation process. 
It is such a relation that was found in NGC 3766 by \citet{mow16}, with which we will compare our model predictions in the next section.
We also note that the PL relations might be blurred in older clusters if rotational velocity decreases with time in some stars due to various braking effects.

The period luminosity relation becomes slightly steeper and covers a wider range of luminosities due to the effect of evolution.
This is shown in Fig.\,\ref{fig:PL}, which compares model predictions at ages of $3\times10^7$\,yr and $1\times10^8$\,yr.
We note that the $6\,M_\odot$ model enters the SPB instability region during its evolution, as a result of its luminosity increase with age.

\section{Comparison with observations}

\label{Sect:ngc3766}

\begin{figure*}
	\includegraphics[width=0.8\textwidth]{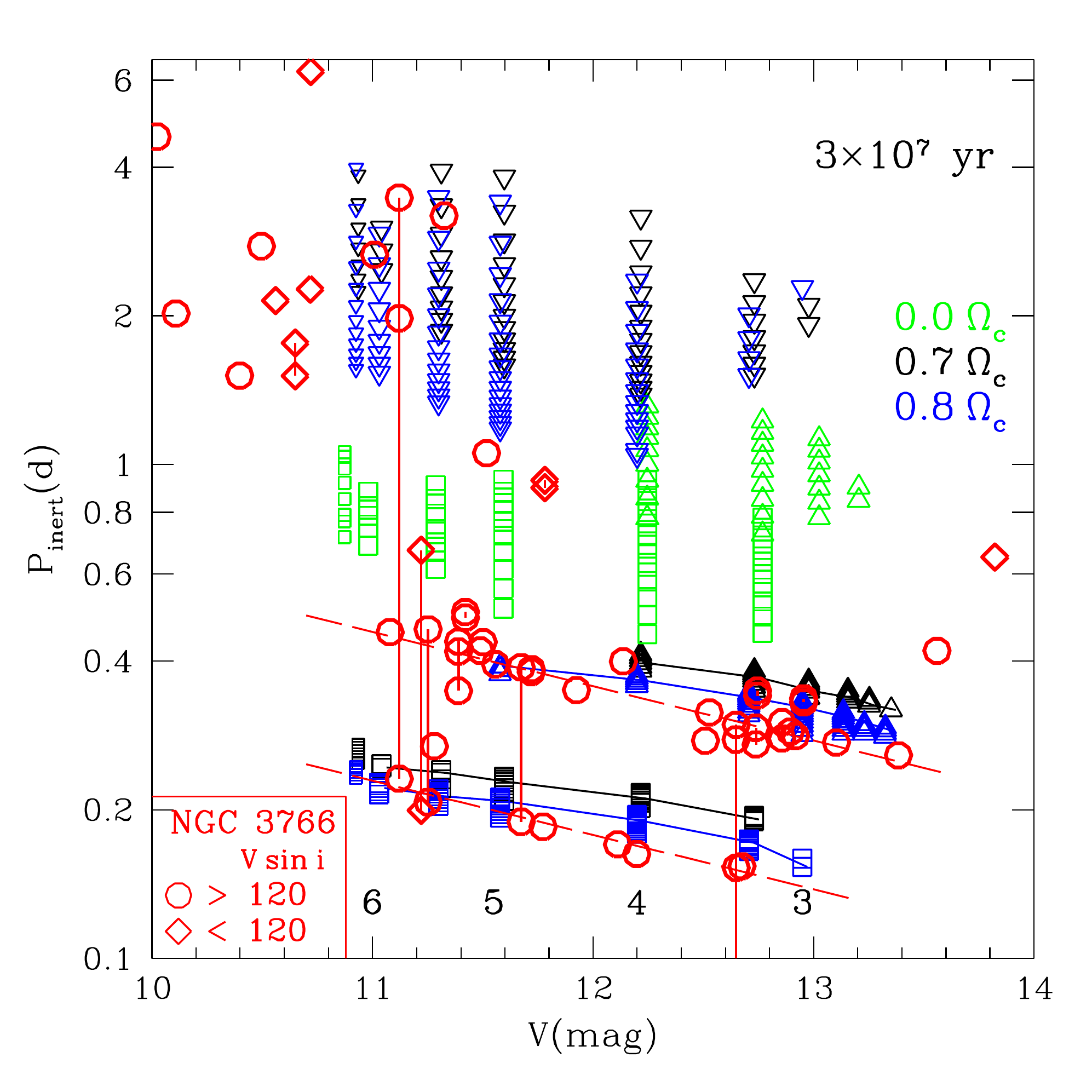}    
        \caption{Period-magnitude diagram.
        Red markers represent the B-type stars observed in NGC~3766 (circles for stars with $v\sin i>120$~km/s and diamonds for stars with $v\sin i<120$~km/s).
        Green, 
        black and blue markers represent the pulsation periods, in the inertial frame, predicted to be excited in $3\times10^7$yr old models having Z=0.014 and initial angular rotation velocities equal to 0, 
        0.7 and 0.8 times the critical angular velocity, respectively.
        Triangles and squares represent excited prograde sectoral g-modes of $m=-1$ and $m=-2$, respectively (with the convention that negative $m$ values correspond to prograde modes), while inverted triangles represent excited $r$~modes of $m=1$.
        Model luminosities are converted to apparent $V$ magnitudes using a distance modulus of $11.61$\,mag, a mean color excess of $E(B-V)=0.22$\,mag \citep{mcs08,aid12}, and bolometric corrections from \citet{flo96}.
        The continuous lines connect the most strongly excited prograde sectoral $g$ modes.
        The dashed lines indicate the two sequences of the PL relation obtained by \citet{mow16} for the FaRPB stars in NGC~3766.
        The numbers 6, 5, 4, and 3 aligned horizontally at the bottom of the panel near the x-axis indicate model masses in solar units.
        The periods of excited modes predicted in $6\,M_\odot$ models with $Z=0.02$ are also shown for comparison. For clarity, they are reported in the figure using smaller symbols at locations slightly shifted leftward from the standard $6\,M_\odot$ models. 
        }
\label{fig:PVmag}
\end{figure*}

\begin{figure}
    \includegraphics[width=\columnwidth]{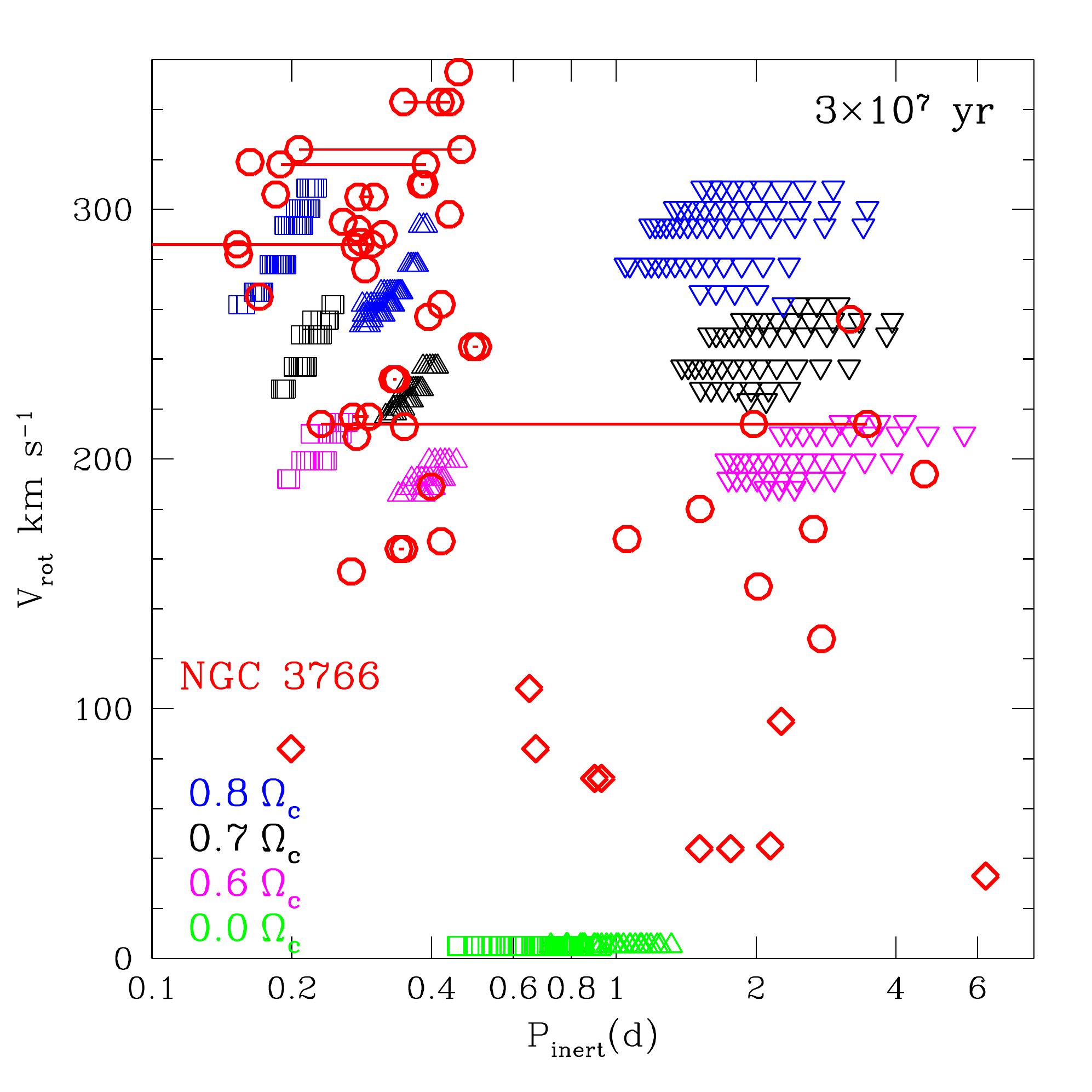}    
    \caption{Projected rotation velocity versus photometric period for observational quantities, or rotation velocity versus predicted excited pulsation periods for quantities derived from stellar models (model rotation velocities are calculated using mean stellar radii and rotation frequencies).
             The meaning of the symbols are the same as in Fig.\,\ref{fig:PVmag}.
             Solid red lines connect the photometric periods of individual multi-periodic stars.
    }
    \label{fig:PVel_ngc3766}
\end{figure}

Our PL relation predictions for fast-rotating stars are compared in Fig.~\ref{fig:PVmag} to the PL relations of FaRPB stars discovered by \citet{mow16} in NGC~3766 (shown as dashed lines in Fig.~\ref{fig:PVmag}).
We take the models at an age of $3\times10^7$~yr (black lines in Fig.~\ref{fig:PL}), in agreement with the age estimate of $30.7\pm7.9$~Myr derived by \citet{aid12}.
A distance of 2.1~kpc has moreover been used to convert the luminosities of our models to apparent $V$ magnitudes.

The sequences of the prograde dipole ($-m = \ell_{\rm e} = 1$) and quadrupole ($-m=\ell_{\rm e} = 2$) modes of rapidly rotating models ($\Omega_\mathrm{ini}/\Omega_\mathrm{c} \gtrsim 0.7$) roughly agree with the observed two sequences of the PL relation.
The predictions of all the fast-rotating models we studied nicely cover the magnitude range of FaRPB stars observed on the second (shortest periods) sequence.
The match in magnitude is less complete for the first (longest periods) sequence, though, the predictions of the fastest-rotating models (with $\Omega_\mathrm{ini}/\Omega_\mathrm{c} = 0.8$) covering only 3/4 of the magnitude range.
To reproduce the brightest FaRPB stars on the first sequence, at $V \lesssim 11.6$~mag, the dipole $g$-modes must be excited in models up to $\sim6\,M_\odot$.
Those modes are, however, damped in our models for $M > 5\,M_\odot$.

Two options can help to solve this discrepancy between observed luminosities of the brightest FaRPB stars and model predictions.
The first option is to assume a larger-than-solar metallicity for NGC~3766.
This is probable because the metallicity of NGC~3766  has been estimated only by a photometric method from the mean ultraviolet excess of bright dwarfs \citep{tad03}, which has a large uncertainty.
We expect that a larger metallicity yields a higher Fe-opacity bump and hence stronger excitation.
To quantify the impact of a higher-than-solar metallicity on the dipole $g$-mode excitation, $6\,M_\odot$ models with a metallicity of $Z=0.02$ have been computed, and their pulsation properties analyzed.
The periods of the excited $g$ and $r$ modes in these $Z=0.02$ models are shown in Fig.~\ref{fig:PVmag}  by smaller symbols at locations slightly shifted leftward from the standard $6\,M_\odot$ models. 
The number of $r$ modes ($m=1$) and $g$ modes of $\ell_{\rm e}=2$ that are excited is larger in the $Z=0.02$ models than in the Z=0.014 models;i.e., a larger metallicity helps excite more modes.  
However, dipole $g$ modes in the $6\,M_\odot$ models, which are needed to explain the most luminous FaRPB stars, are still not excited even in the most rapidly rotating model, which indicates that the enhancement of metallicity with the OPAL opacity is insufficient to increase the mass limit for the excitation of dipole $g$ modes to $6\,M_\odot$.

The second option that can help to excite dipole $g$ modes in stars more massive than $5\,M_\odot$ is to use OP \citep{bad05} rather than OPAL \citep{igl96} opacities.
Test studies using OP opacities on non-rotating models show that the SPB instability range is shifted to slightly higher effective temperatures than the ones obtained using OPAL opacities (see Appendix~\ref{sec:opalop} and Fig.~\ref{fig:opalop}).
A similar conclusion is reached by \citet{mig07}.
The opacity effect would be stronger \citep{sal12,mor16L,das17} if the Ni and possibly Fe opacities are revised upward compared to OP opacities as discussed by \citet{tur16}.  
The combined effect of the two factors, higher cluster metallicity and use of OP opacities, may then act to increase the stellar mass up to which dipole $g$~modes can be excited 
 (as shown by the blue solid line in Fig.~\ref{fig:opalop}),
thereby improving the match between model predictions and observations of FaRBP stars that obey the newly discovered PL relation in NGC~3766.

We note that the gradient of the observational PL relations are slightly steeper than that along a constant $\Omega_{\rm init}/\Omega_{\rm c}$, making the predicted periods slightly longer than the observed ones in the less luminous part. 
This may indicate less massive stars to be rotating more rapidly, or a higher metallicity/opacity as discussed above may reduce the discrepancy.
Furthermore, although beyond the scope of this paper, a careful statistical comparison between observed PL relations and theoretical ones based on models with improved metallicity/opacity would yield a mean $M-\Omega$ relation along the main sequence of the cluster.

In addition to the FaRPB stars satisfying the PL relation, periodic variable stars were observed in NGC~3766 with periods larger than 1 day \citep{mow13} and in the same magnitude range as the FaRPB stars.
\citet{mow16} showed that several of them could be explained by the fact that they are in binary systems, with the photometric period linked to their orbital period.
But this is not the case for all of them.
Interestingly, one of the single stars \citep[having star ID 51 in][]{mow13} contains both a short (0.23111~d, falling on the second PL relation) and long (3.4692~d, 1.9777~d) periods.
The two longer periods are much too long to be explained by $g$-mode pulsations, but they are comparable to the periods predicted for excited $r$~modes (see Fig.\,\ref{fig:PVmag}, where the $r$~modes are drawn with inverted triangles).
The star is rotating at $v\sin i= 214 \pm 21$~km/s \citep{mow16}, which is compatible with the requirement of fast rotation to generate $g$ and $r$~mode pulsations.
The fact that both short and long periods are simultaneously excited in the same star further supports an $r$-mode pulsation for the long periods in this star.

The rotation velocities of the $3\times10^7$\,yr models that lead to excited $g$ and $r$~modes are shown in Fig.\,\ref{fig:PVel_ngc3766} versus the predicted pulsation periods.
The rotation velocities are calculated based on the rotation frequencies and mean radii of the stellar models.
The equatorial velocities should actually be slightly larger than those calculated using mean stellar radii due to the oblateness of fast-rotating stars, but by less than a few percents. 
The observed $v\!\sin\!i$ versus period distribution of FaRPB stars in NGC\,3766, plotted with red circles and diamonds in Fig.\,\ref{fig:PVel_ngc3766}, shows good agreement with our model predictions: the majority of stars whose periods fall on either or both PL sequences have sufficiently high $v\sin i$ values to be consistent with our explanation that these sequences are formed by prograde sectoral $g$ modes of $m=-1$ and $m=-2$ in rapidly rotating stars.
Such prograde sectoral modes have pulsation amplitudes that are maximum at the equator and that decrease with decreasing inclination angle.
FaRPB stars are thus seen close to equator-on, and their equatorial rotation velocity should be close to (yet statistically larger than) the measured $v \sin i$.
For the $\ell_{\rm e}=1, m=-1$ mode, for example, the pulsation amplitude is predicted to reach half the maximum amplitude at $i \simeq 53^\mathrm{o}$  
\citep[see Fig.~2 of][]{sal14}, and the equatorial rotation velocity would be 25\% larger than the measured $v \sin i$.
The fact that no FaRPB star is observed with a low $v \sin i$ further supports a prograde sectoral $g$-mode origin for their pulsation periods.

\begin{figure}
    \includegraphics[width=\columnwidth]{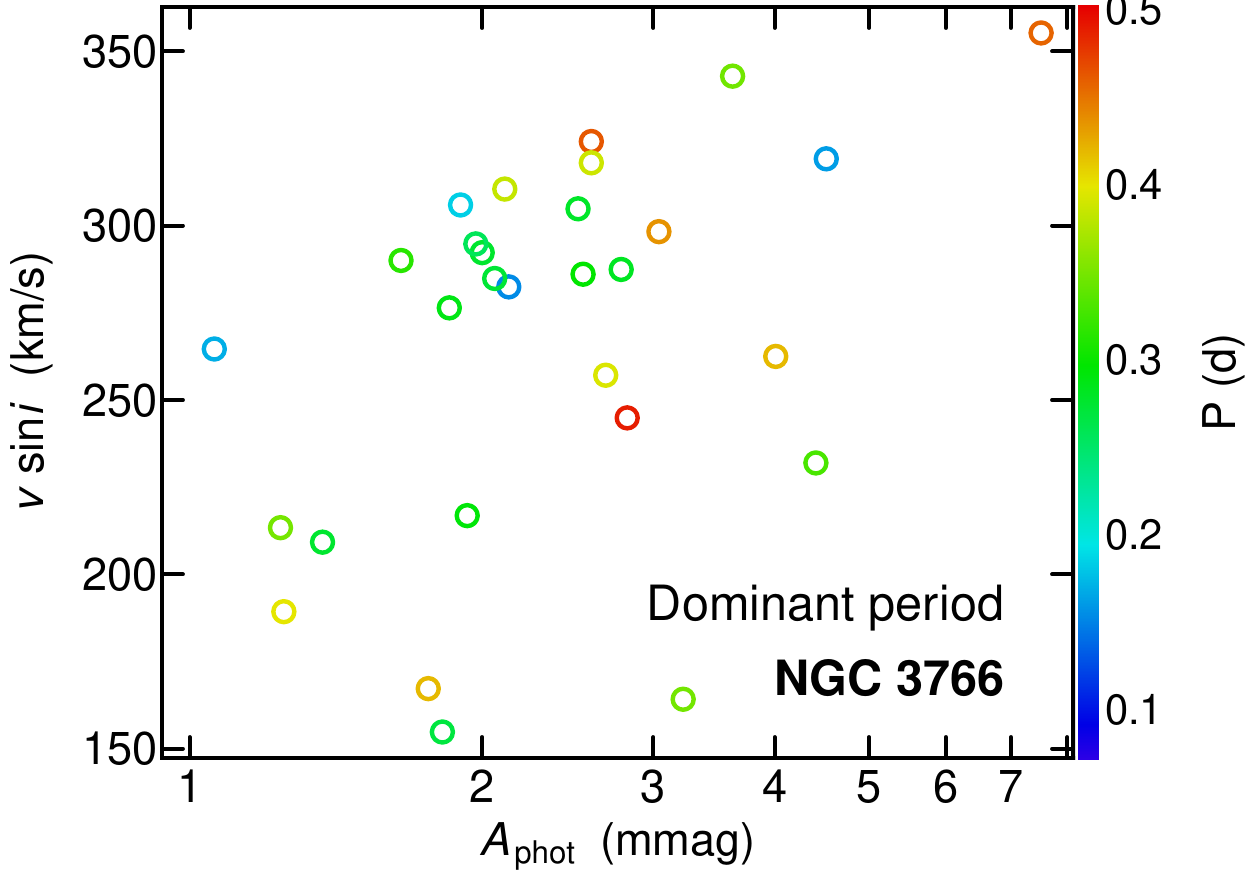}      
    \caption{Projected rotational velocity $v\sin i$ versus photometric variability amplitude $A_\mathrm{phot}$ of all FaRPB stars with photometric periods smaller than 0.55~d.
    Note that the horizontal axis is scaled logarithmically. 
    Only the dominant period (i.e. with the largest variability amplitude) is plotted for multi-periodic stars.
    The period is shown in color according to the color scale drawn on the right of the figure.
    }
    \label{fig:vsiniVsAmplitude}
\end{figure}

 The observed distribution of the photometric amplitude $A_\mathrm{phot}$ of the dominant period of variability versus $v \sin i$ also supports a dipole $g$-mode origin of the pulsation. This distribution is shown in Fig.~\ref{fig:vsiniVsAmplitude} for all FaRPB stars with periods less than 0.55~d.
An upper envelope is visible in the figure, showing a lack of small-amplitude variable stars with large $v\sin i$.
This is consistent with prograde sectoral $g$ modes having their maximum pulsation amplitude at the equator.
Such pulsating stars would have both a small $v \sin i$ and a small variability amplitude when seen non equator-on.

Pulsating stars with periods longer than $\sim 0.5$~d form a more inhomogeneous group.
While all stars with periods shorter than 0.5~d are fast rotating, long-period pulsators are observed with a variety of $v\!\sin\!i$ values (see Fig.\,\ref{fig:PVel_ngc3766}).
The origin of the long-period variables with small  $v\!\sin\!i$ can be $g$-mode pulsations similar to the classical (i.e. non-rotating) SPB stars.
They are also, on the mean, brighter than the fast-rotating stars (see Fig.~\ref{fig:PVmag}).
The long-period variables with large $v\!\sin\!i$, on the other hand, can be attributed to $r$-mode pulsations.
Among them is the fast-rotating hybrid star having both $g$- and $r$- mode pulsation frequencies.
The distinction between $g$- and $r$-mode pulsations would be apparent if period spacings could be obtained from future observations (cf. Fig.\,\ref{fig:space}).

\section{Conclusions}
\label{Sect:conclusions}

We found that the PL relations of short-period ($\lesssim 0.5$\,d) FaRPB stars discovered by \citet{mow16} in NGC\,3766 can be explained by $m=-1$ and $m=-2$ prograde sectoral $g$ modes of rapidly rotating stars born with initial angular velocities larger than about 70\,\% of the critical angular velocities.

Such PL relations would be blurred in older clusters if the rotation of B-type stars is slowed down by such mechanisms as, for example, magnetic breaking,  
tidal effects in binary stars, or
possibly angular momentum transport by internal gravity waves \citep{rog15}.
It would be interesting to obtain the PL relations in open clusters of various ages. This may provide a way to obtain information on the rotation slow-down timescale of B-type stars.

Some rapidly rotating B-type stars in NGC~3766 also show long period variations ($\ga 1$\,d). We showed that some of them can be explained as stars pulsating in $r$ modes excited by the same kappa-mechanism as $g$ modes.

\section*{Acknowledgements}
We thank the anonymous referee of this paper for useful comments.
HS thanks George Meynet for useful discussions and his hospitality in Geneva Observatory, 
and Umin Lee for helpful discussions.


\bibliographystyle{mnras}
\bibliography{ref}

%
\appendix
\section{Eigenvalue of Laplace's Tidal equation}\label{sec:lambda}
\begin{figure}
\includegraphics[width=\columnwidth]{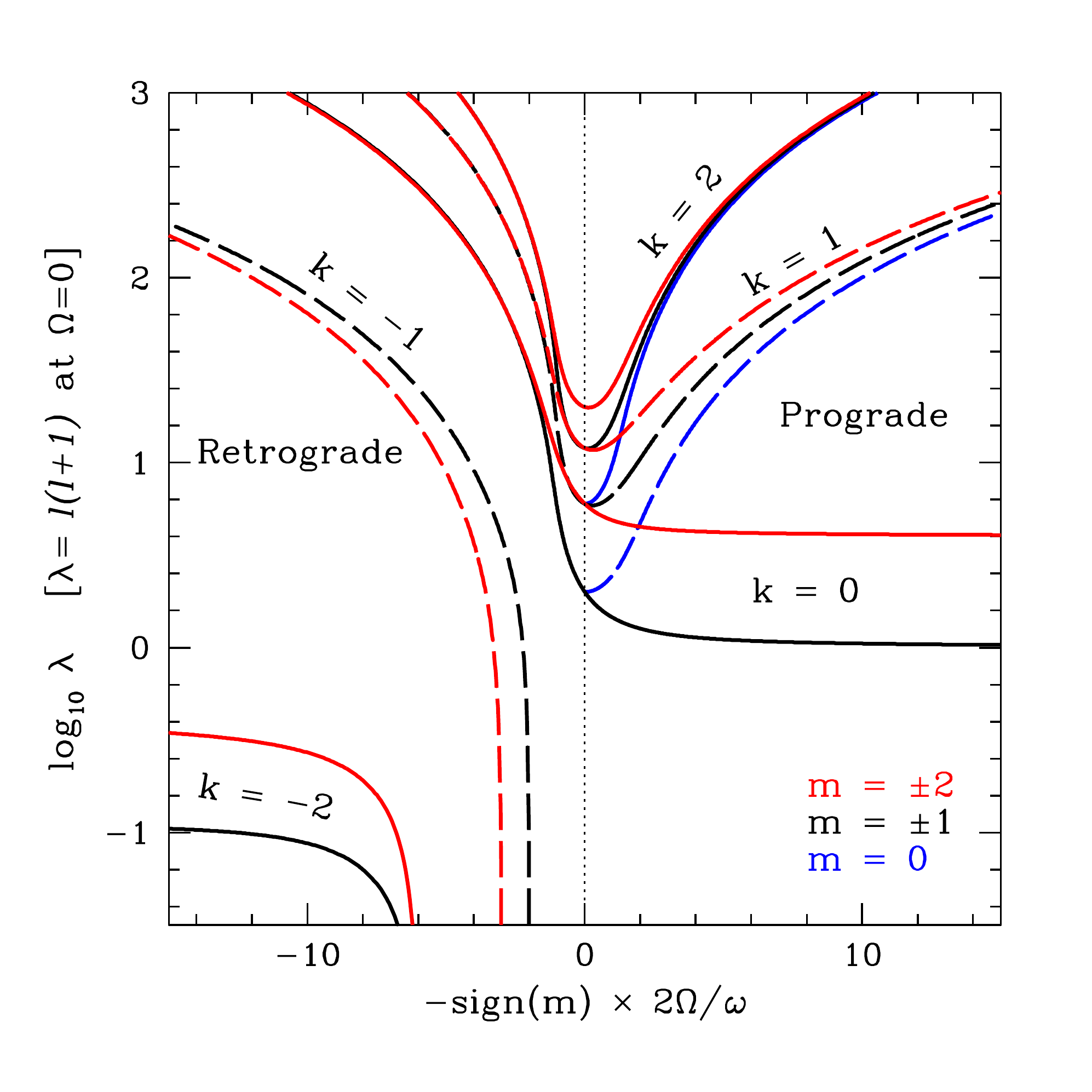}      
\caption{ The eigenvalue of Laplace's tidal equation $\lambda$ as a function of the spin parameter $2\Omega/\omega$, where $\omega$ is the angular frequency of pulsation in the co-rotating frame.
The value of azimuthal order $m$ is color-coded as indicated. Solid and dashed lines are used for even and odd modes (with respect to the equator), respectively. The way of ordering $\lambda$ by integers $k$ is adopted from \citet{lee97}. 
}
\label{fig:lambda}
\end{figure}

Although we do not use the traditional approximation of rotation (TAR) in our linear nonadiabatic analysis in this paper, the TAR is useful in understanding the property of low-frequency oscillations of a rotating star.  
In this Appendix section we discuss the properties of $g$ and $r$ modes using the TAR.
Under the TAR, the angular dependence of a pulsation mode of a rotating star is separable from the radial dependence.
The angular dependence is obtained by solving Laplace's tidal equation \citep[e.g.,][and references therein]{lee97} with an eigenvalue $\lambda$, which represents the latitudinal degree of the equatorial concentration; a larger $\lambda$ means the eigenfunction to be more strongly concentrated toward the equator.  
The value of $\lambda$ for a given $m$ depends on the spin parameter $2\Omega/\omega$, where $\Omega$ is the rotation frequency, and $\omega$ is the pulsation frequency in the co-rotating frame.
Fig.\,\ref{fig:lambda} shows $\lambda$ as a function of the spin parameter for the cases of  azimuthal orders of $m=0$ (blue), $\pm1$ (black), and $\pm2$ (red). 
The $\lambda$ is ordered by an integer $k$ as in \citet{lee97}, in which 
the $g$ modes and the $r$ modes correspond to $k \ge 0$ and $k \le -1$, respectively. 

First, we discuss the properties of $g$ modes ($k \ge 0$).
As $\Omega \rightarrow 0$,  $\lambda$ reduces to $\ell(\ell+1)$ with $\ell = |m| + k$, where $\ell$ is the latitudinal degree of spherical harmonics $Y_\ell^m(\theta,\phi)$.
Thus, the sectoral modes correspond to $k=0$, and even and odd tesseral modes correspond to even and odd $k$, respectively.  
As seen in Fig.\,\ref{fig:lambda}, except for the prograde sectoral (i.e., $m < 0$ and $k = 0$) modes, $\lambda$ becomes very large as $2\Omega/\omega$ increases, indicating the eigenfunctions to become  concentrated to the equator as the spin parameter increases.
(This means that in our numerical calculations with expanding eigenfunctions as eqs.~\ref{eq:expand1},~\ref{eq:expand2}, terms associated with larger $l_j$ become important.)
For this reason, all the tesseral $g$ modes and retrograde sectoral ($m > 0, ~k = 0$) $g$ modes are expected to be invisible if $2\Omega/\omega > 2$, which is consistent with the result of \citet{tow03} (see also Appendix~\ref{sec:growth}).   
On the other hand, for the prograde ($m < 0$) sectoral ($k=0$) $g$ modes, $\lambda$ decreases and approaches $m^2$ as $2\Omega/\omega$ increases; i.e., the prograde sectoral $g$ modes are least concentrated toward the equator and most visible among the $g$ modes for a given $m$ in a rapidly rotating star.

The $r$ modes correspond to $\lambda$ with a negative $k$.   
They are always retrograde modes ($m > 0$) in the co-rotating frame.
The $r$ modes can exist in the range of spin parameter where  $\lambda$ is positive; i.e., $2m\Omega/\omega > (|m|+|k+1|)(|m|+|k|)$. 
In other words, the frequencies of $r$ modes are bounded as 
\begin{equation}
\omega_{r {\rm mode}} < {2m\Omega\over (m + |k+1|)(m+|k|)} \quad  \le \,\Omega  
\end{equation}
with $m \ge 1$ and $k \le -1$.
(The last inequality indicates that the $r$ modes are seen as prograde modes in the inertial frame.)
At the limiting frequency, $\lambda = 0$, which corresponds to the fact that oscillation velocities are purely toroidal with diminishing horizontal divergence; i.e., $\nabla_{\rm h}\cdot\bm{\eta} = 0$ (see eqs.~\ref{eq:nabla} and \ref{eq:toroid}).
(The horizontal divergence of a $g$ mode at $\Omega=0$ is proportional to $\ell(\ell+1)$.)
The $r$ modes are characterized as retrograde modes (in the co-rotating frame) dominated by toroidal motions, which corresponds to small $\lambda$.
However, the modes of $k=-1$ become to have $g$-mode character with large $\lambda$ as $2\Omega/\omega$ increases, while the $r$-mode character is retained for the modes with $k\le -2$.
The amplitude of a typical $r$ mode is broadly confined to the mid-latitude zone \citep[e.g.,][; see also Appendix~\ref{sec:growth}]{lee97,sav05}, so that $r$ modes can be photometrically visible (in particular, longer period ones in the inertial frame) if sufficiently large temperature variations are generated.

\section{Growth rates and amplitude distribution on the surface}\label{sec:growth}
\begin{figure}
\includegraphics[width=\columnwidth]{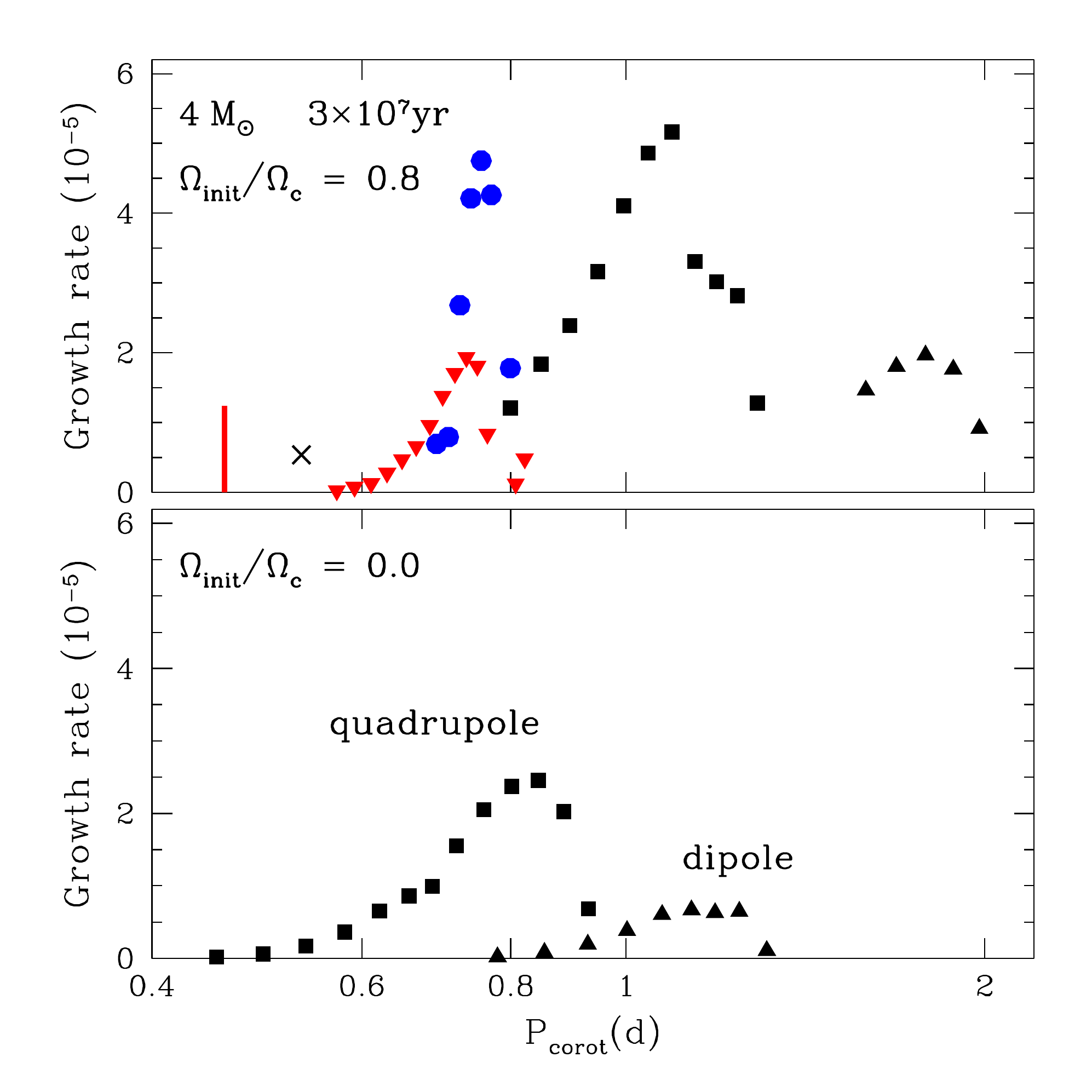}        
\caption{Growth rates of excited modes versus the periods in the co-rotating frame in the non-rotating (botom panel) and rotating (top panel) models of $4\,M_\odot$ at an age of $3\times10^7$yr. 
Filled black triangles and squares are prograde sectoral $m=-1$ and $m=-2$ ($\ell=1$ and $\ell=2$ in the non-rotating model) modes, respectively.
A cross is for a single prograde tesseral mode of $(m,\ell_{\rm e})= (-1,2)$ excited in the rotating model.
Filled blue circles are axisymmetric ($m=0$) odd tesseral modes of $\ell_{\rm e} = 1$. 
Red inverted triangles are excited odd $r$ modes of $m=1$.
The short vertical bar indicates the rotation period.  
}
\label{fig:growth}
\end{figure}
Fig.~\ref{fig:growth} shows growth rates versus periods in the co-rotating frame for excited modes of $|m|\le 2$ in a rotating $4\,M_\odot$ model with $\Omega_{\rm init}=0.8\Omega_{\rm c}$ (top panel) and a non-rotating $4\,M_\odot$ model (botom panel) at an age of $3\times10^{7}$\,yr.
The number of prograde sectoral $g$ modes excited in the rotating star is comparable to the number of excited $\ell=1$ and 2 $g$-modes in the non-rotating model, while only one prograde tesseral, $(m,\ell_{\rm e}) = (-1,2)$, $g$ mode ($\times$) is excited in the rotating model.
A few axisymmetric odd modes ($m=0, \ell_{\rm e}=1$) are excited, while all the even axisymmetric modes of $\ell_{\rm e} = 2$ are found to be damped.
Since the period in the co-rotating frame of all the excited tesseral $g$ modes are longer than the rotation period, the corresponding spin parameters are greater than 2.

We note that some $m=0$ even modes with periods of $\sim 0.2$~d ($2\Omega/\omega < 1$) corresponding to high $\ell$ ($\approx 6$) are excited. Those modes are not shown here, because they should have low visibilities. \citet{bal99} first found such $g$ modes of high $\ell$s to be excited in B-type main-sequence stars.

\begin{figure}
\includegraphics[width=\columnwidth]{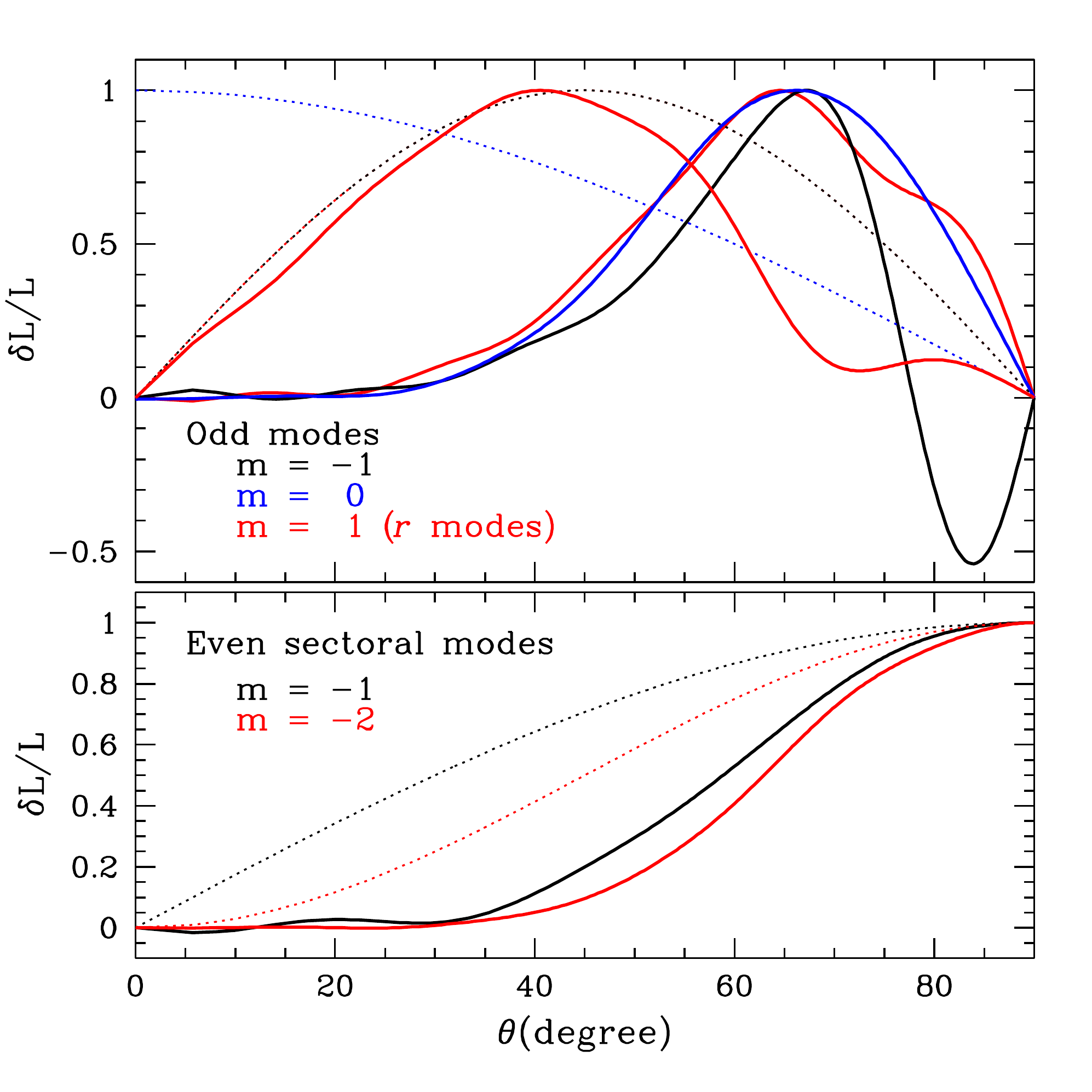}       
\caption{Amplitude of local luminosity variation (i.e., linear variation of $R^2T^4$) versus polar-angle (co-latitude) for selected modes shown in Fig.~\ref{fig:growth}. Upper and lower panels are for odd and even modes, respectively. 
Each distribution is normalized as unity at the maximum.
Corresponding azimuthal order $m$ is color-corded as indicated.
Dotted lines show amplitude distributions expected in a non-rotating case;i.e., Legendre function $P_\ell^{|m|}(\cos\theta)$. 
Amplitude distributions for two $r$ modes are plotted (red lines in the top panel); one at $P_{\rm corot}=0.567$\,d and the other at $P_{\rm corot}=0.734$\,d.  The former having a broad peak around $\theta \approx 40^\circ$ is the lowest radial order ($n=7$) $r$ mode exited, and the latter ($n=17$) with the maximum growth rate.
The amplitude distributions in the bottom panel are for the sectoral prograde modes having maximum growth rates. 
}
\label{fig:amp}
\end{figure}
Fig.~\ref{fig:amp} shows the amplitude distributions of luminosity variation on the stellar surface of selected modes from Fig.~\ref{fig:growth}.
The amplitude distributions of the odd tesseral $g$-modes (blue and black solid lines in the upper panel) are significantly modified from the corresponding Legendre functions, and shifted toward the equator ($\theta = 90^\circ$) by the effect of rotation.
We do not expect to detect any of these tesseral odd modes in stars having large values of $v\sin i$, because of cancellation.
In contrast to the tesseral $g$ modes, rotational effects on the prograde sectoral modes are gentle (solid lines in the bottom panel of Fig.~\ref{fig:amp});  the amplitude distribution is shifted toward the equator without changing the shape significantly.
We expect these sectoral prograde modes to be most visible in stars with large $v\sin i$.

All excited retrograde modes are found to be $r$ modes (inverted red triangles in Fig.\,\ref{fig:growth} and red solid lines in the top panel of Fig.~\ref{fig:amp}); i.e. no retrograde $g$ modes are excited.  
These $r$ modes are odd modes corresponding to the sequence of $k=-1 ~(m=1)$ in Fig.~\ref{fig:lambda}.  
The $\lambda$ of the sequence increases rapidly as the spin parameter ($2\Omega/\omega$) increases (so does $P_{\rm corot}$), and hence the property shifts from a pure $r$ mode (with $\ell_{\rm e}=2$) to that mixed with the $g$ mode property. This means that components associated with $\ell_j > 2$ become important in the expansions of eqs.~\ref{eq:expand1},~\ref{eq:expand2} as $2\Omega/\omega$ increases.
Correspondingly, the distribution of amplitude,  in Fig.~\ref{fig:amp} (red lines in the top panel), shifts from that close to the Legendre function $P_2^1$ for the shortest period (excited) $r$ mode ($P_{\rm corot}=0.567$\,d) to that concentrated toward the equator (gets  similar to tesseral $g$ modes) for the mode at $P_{\rm corot}=0.734$\,d (with a maximum growth rate).
In other words, the visibility of odd $r$ modes would decrease as $P_{\rm corot}$ increases (i.e., as $P_{\rm inert}$ {\it decreases}).

\section{SPB instability ranges obtained by OPAL and OP opacities}
\label{sec:opalop}
\begin{figure}
\includegraphics[width=\columnwidth]{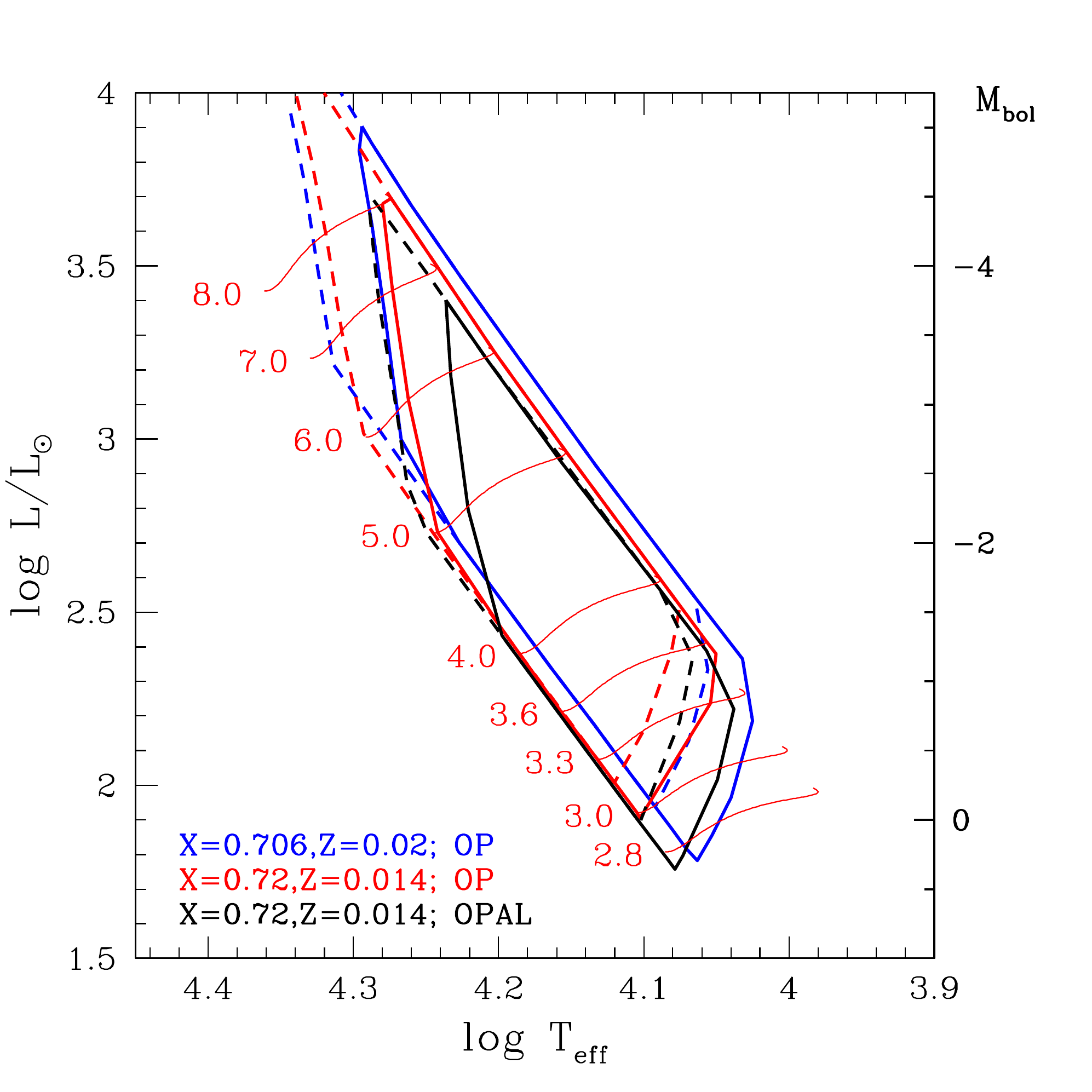}           
\caption{Theoretical HR diagram showing the instability boundaries of $g$ modes for non-rotating models obtained using the OP and OPAL opacities.
Corresponding chemical compositions and opacity tables are color coded as indicated.
Solid and dashed lines are for $\ell =1$ and $\ell = 2$ modes, respectively.
The thin red lines represent evolutionary tracks of models computed with the MESA code.
The numbers along the ZAMS indicate the stellar mass of the track, in solar units.
}
\label{fig:opalop}
\end{figure}

In the Geneva evolution code OPAL opacity tables \citep{igl96} are used.
To see the difference in the excitation of $g$ modes between OPAL  and OP \citep{bad05} opacities, we have also analysed the excitation of $g$ modes for non-rotating models obtained by the Modules of Experiments in Stellar Astrophysics \citep[MESA][]{pax13} using the OP opacity.
Fig.\,\ref{fig:opalop} shows the difference.
In accordance with \citet{mig07}, the SPB instability range obtained  by the OP opacity is shifted slightly to higher $T_{\rm eff}$ compared with the case of the OPAL opacity. 
Furthermore, if the OP opacity is used with a slightly higher metallicity of $Z=0.02$ (the blue lines in Fig.\,\ref{fig:opalop}), the instability region also extends to lower luminosity, covering the instability regions obtained with OPAL and OP opacities for $Z=0.014$.
Thus, combining a slightly higher-than-solar metallicity and the OP opacity  
would solve the lack of excitation in the luminous part of Fig.\,\ref{fig:PVmag}. 


\bsp	
\label{lastpage}
\end{document}